%% file: tmplans.tex
\documentclass[graybox]{svmult}
\usepackage{url}
\usepackage{graphicx}
\usepackage{amsmath,amssymb}

\usepackage{multirow}
\usepackage{mathptmx}       % selects Times Roman as basic font
\usepackage{helvet}         % selects Helvetica as sans-serif font
\usepackage{courier}        % selects Courier as typewriter font
\usepackage{type1cm}        % activate if the above 3 fonts are
\usepackage[bottom]{footmisc}% places footnotes at page bottom

\date{}

\begin{document}

\title*{Plan Interdiction Games}
\author{Yevgeniy Vorobeychik and Michael Pritchard}
\institute{Yevgeniy Vorobeychik \at Department of Electrical Engineering and
Computer Science, Vanderbilt University, Nashville,
  TN \email{eug.vorobey@gmail.com}
\and
Michael Pritchard \at Department of Electrical Engineering and
Computer Science, Vanderbilt University, Nashville,
  TN \email{michael.j.pritchard@vanderbilt.edu}}

\maketitle

%\abstract*{TODO}

\abstract{
%Cyber risk assessment and mitigation are major issues faced by
%  organizations.  
We propose a framework for cyber risk assessment and
  mitigation which models attackers as formal planners and defenders
  as interdicting such plans.  We illustrate the value of plan
  interdiction problems by first modeling network cyber risk through
  the use of formal planning, and subsequently formalizing an
  important question of prioritizing vulnerabilities for patching in
  the plan interdiction framework.  In particular, we show that 
  selectively patching relatively few vulnerabilities allows a network
  administrator to significantly reduce exposure to cyber risk.
  More broadly, we have developed a number of
  scalable approaches for plan interdiction problems, making
  especially significant advances when attack plans involve
  uncertainty about system dynamics.  However, important open problems
remain, including how to effectively capture information asymmetry
between the attacker and defender, how to best model dynamics in the
attacker-defender interaction, and how to develop scalable algorithms
for solving associated plan interdiction games.}

\input{intro}
\input{example}
\input{planning}
\input{planinterdiction}
\input{illustration}
\input{dynamicpi}
\input{conclusion}

\bibliographystyle{spmpsci}
\bibliography{plan,references}

\end{document}

%% file: intro.tex
\section{Introduction}

Interdiction seems by its very nature an adversarial act, one perpetrated, if you
will, by ``bad guys.''
For example, an attacker may interdict the power flow on an electric
power grid, resulting in widespread blackouts~\cite{Salmeron09}, or
interdict a transportation or a supply
network~\cite{McMasters70,Ghare71}.
Consequently, it may be somewhat jarring at first to consider the
\emph{defender} ---the ``good guy''--- as the interdictor.
We would like to argue that in cybersecurity this is precisely the
ultimate goal of the defender: to \emph{interdict} an attack, whether
actual or potential.
In particular, in this chapter we will illustrate that cyber attacks
are naturally viewed as plans, or dynamic decision processes by
malicious agents.

At the most basic level, a plan is a sequence of actions which, if
successfully executed beginning from an appropriate initial state,
accomplishes a planner's goal.
Of course, this notion of a plan is quite restricted; for example, if
uncertainty is at all salient, an intelligent plan would involve
\emph{contingencies}, or a mapping from observed state to action.
The perspective we will take in this chapter is that the \emph{attacker is
  such a planner}.
The initial state for the attacker includes the initial
vulnerabilities of the target network, as well as any relevant
capabilities and information that the attacker possesses.
An attack is then a series of actions ---perhaps, as a part of a
contingent plan of actions--- aimed at accomplishing a malicious goal,
or perhaps a contingent series of goals, each with a different value
to the attacker.
The defender, as we had remarked, is the \emph{interdictor}: insofar as
the attacker's plan would accomplish goals which are contrary to the
defender's desires, the defender would wish to prevent this from
happening.
A myopic defender would simply attempt to interdict past attacks;
arguably, that is what most cyber-defense is like in practice: preventing
attacks which have been identified from succeeding.
We argue for a longer, proactive view: the defender can, and should,
reason about alternative attack plans that the attacker \emph{could
  make}, for particular defensive interdiction strategies, and choose
an \emph{optimal} interdiction ---one that proactively accounts not
merely for previously known attacks, but possible future variations
which circumvent the defense as well.

Below, we begin with a simple example to illustrate the nature of
attack plans and interdictions.
Subsequently, we describe several formal modeling approaches for plan
interdiction, first when the planning environment is deterministic,
and subsequently when it is stochastic.
We follow with an in-depth illustration of the approach by using it as
a modeling framework for network cyber risk assessment.
We then suggest future research directions that consider longer term
attack plans, when the defender is able to observe, and react to, some
of the attack actions.
These directions present interesting questions about optimal adaptive
defense, as well as important considerations of adversarial deception.

%% file: example.tex
\section{A Simple Example}

\begin{figure}[ht]
\centering
\includegraphics[width=4in]{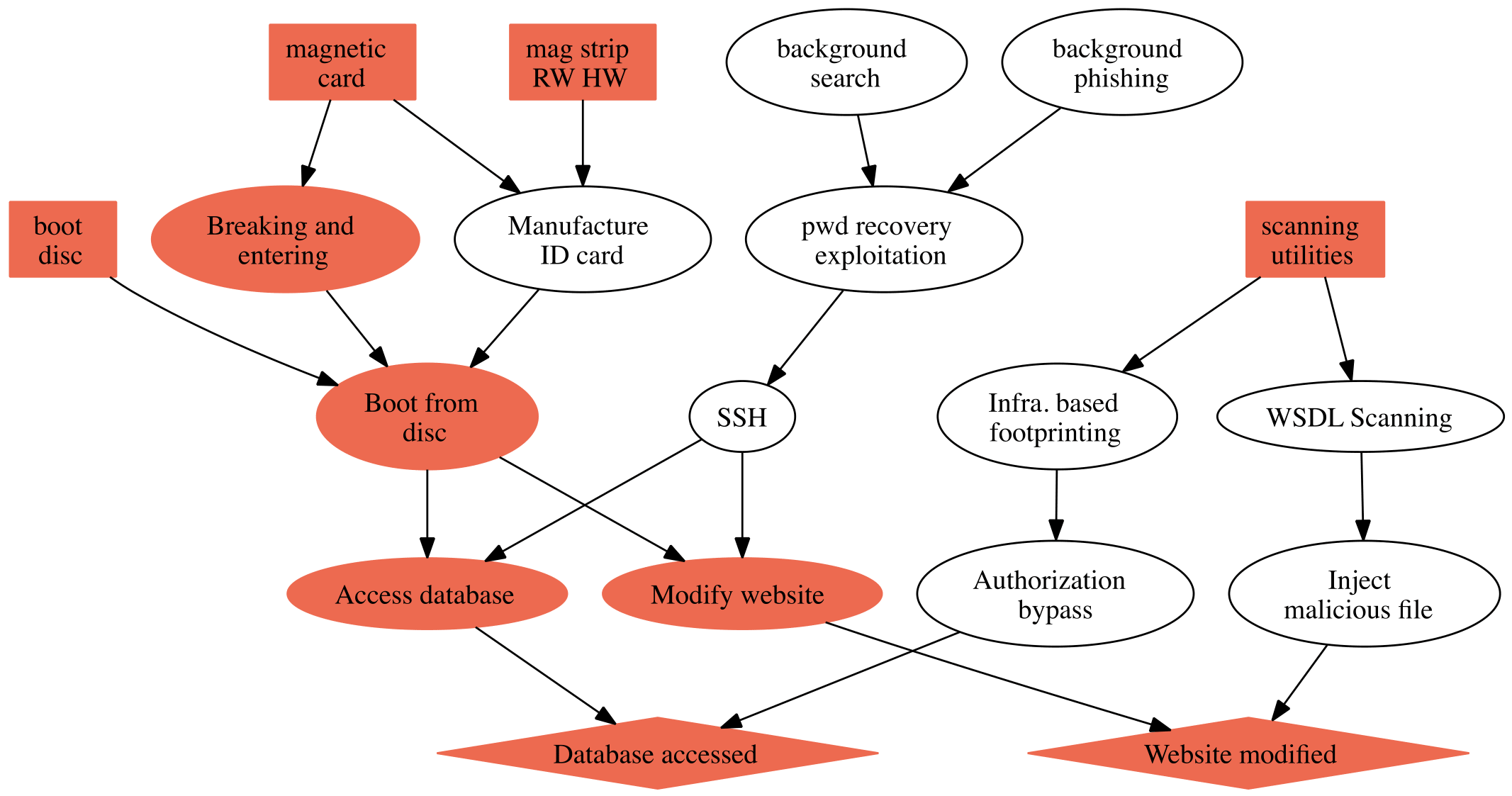}
\caption{Example attack graph. Boxes correspond to initial attacker capabilities, ovals are
attack actions, and diamonds are attacker goals. An optimal attack
plan is highlighted in red.}
\label{F:example}
\end{figure}

As a simple illustration, consider the following example of an attack
planning problem.
The initial state includes the initial attacker capabilities, such as
possession of a boot disk and port scanning utilities.
Actions include both physical actions (breaking and entering and
booting a machine from disk) as well
as cyber actions, such as performing a port scan to find
vulnerabilities.\footnote{The actions in our example are taken from the CAPEC database (\url{http://capec.mitre.org}).}
Figure~\ref{F:example} shows an attack graph (with attack actions as
nodes) for this scenario, with the actual attack plan highlighted in
red.

\begin{figure}[ht]
\centering
\includegraphics[width=4in]{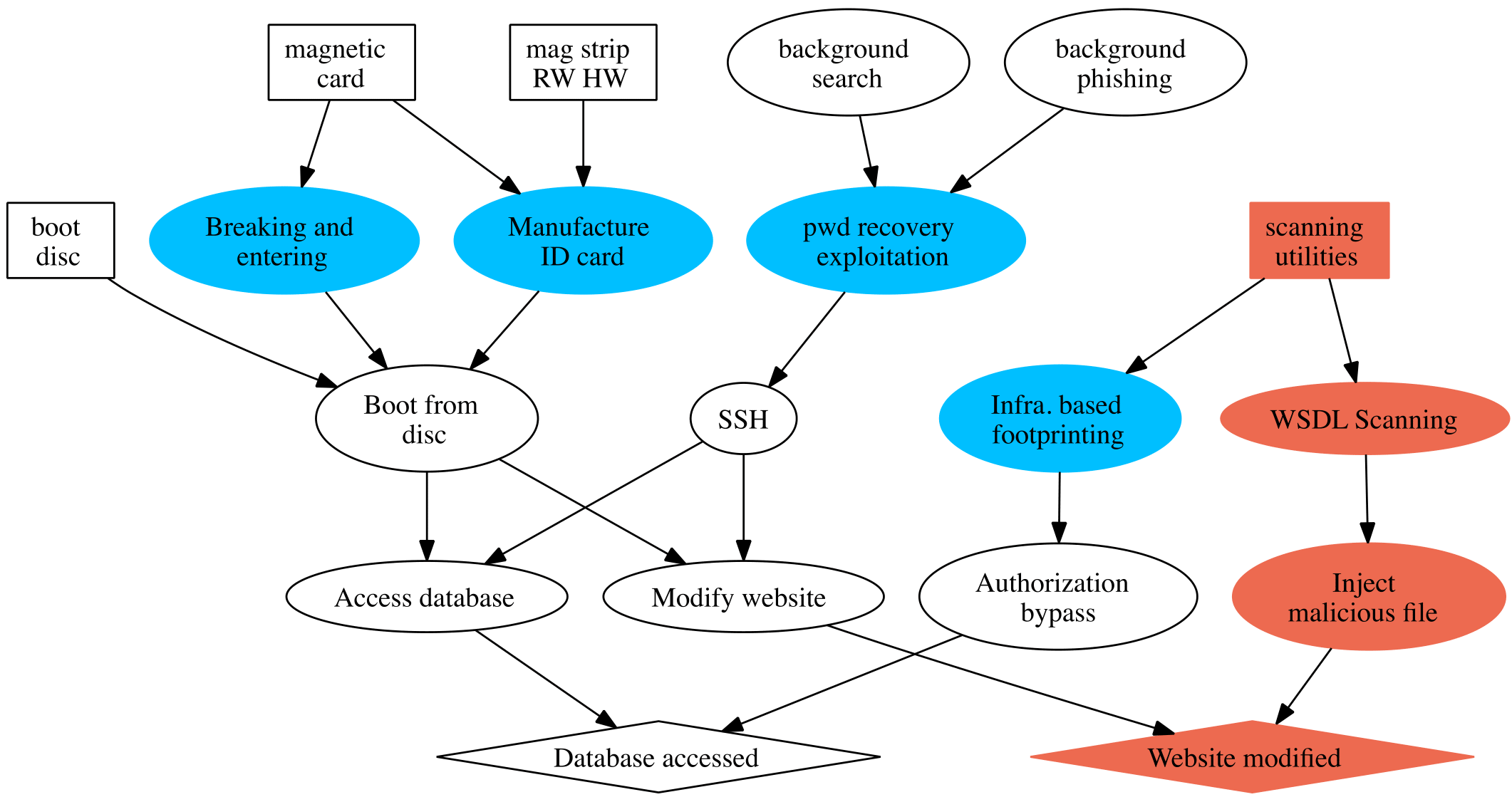}
\caption{Example interdiction plan: actions that are blocked are
  colored in blue, and the final attack plan (circumvention) is
  highlighted in red.}
\label{F:interdictionEx}
\end{figure}

Figure~\ref{F:interdictionEx}, on the other hand, shows an example
interdiction plan.
In this example, we interdict a subset of actions (for example, by
patching specific vulnerabilities, or changing the network architecture).
A noteworthy observation about the particular interdiction strategy in
Figure~\ref{F:interdictionEx} is that it still allows the attacker to
attack a low-stakes target.
This is because, in this specific example, interdicting all
possibilities is not cost-effective, relative to the consequence of
successful attack.

%% file: planning.tex
\section{Planning in AI}

We begin by formalizing the notion of \emph{planning}, starting with
classical planning in deterministic domains, and then considering
generalizations that allow us to capture uncertainty.

\subsection{Classical Planning}

Formally, a classical (deterministic) planning problem is a tuple $\{X,A,s_0,G,r,c\}$,
where $X$ is the set of literals (binary variables) which represent the state of the world relevant for the planning problem, $A$ is
the set of actions, $s_0$ is the set of literals which are initially true
(i.e., the initial state of the world), and $G$ is the set of goals.
A plan action $a \in A$ is characterized by a set of
\emph{preconditions}, that is, the set of literals that must be true
in the current state for the action to be applicable, and a set of
\emph{effects}, which either add literals from current state, or
delete these, thereby determining transition from one state to another.
A reward function $r_l$ assigns a value (utility) to each goal literal
$l \in G$ (we assume that the total utility is additive in these).
Finally, $c_a$ is the cost of taking an action $a$.
A solution to this planning problem is a plan $\pi$, which is a
sequence of actions.
A number of effective approaches exist for solving such planning
problems at scale~\cite{Chen06:Temporal}.

\subsection{Planning under Uncertainty}

A common formalization of planning under uncertainty makes use of
\emph{Markov decision processes (MDPs)}, defined by a tuple
$\{S,A,P,r,p_0\}$,
where $S$ is the set of states (of the world), $A$ the set of actions,
$P(s,a,s') = \Pr\{s'|s,a\}$ a transition function which represents
system dynamics as a function of state $s$ and action $a \in A$ taken
by the planner, $R(s,a)$ the reward function, and $p_0(s)$ the
probability distribution over initial states; below, we assume that
there is a fixed starting state $s_0$~\cite{puterman1994markov}.
A solution to the MDP is a policy, $\pi$, which determines actions to
take as a function of relevant information.
If we further assume that the MDP is discrete-time, infinite-horizon,
and discounted (future is discounted exponentially, with a discount
factor $\gamma \in [0,1)$), we can restrict attention to policies $\pi(s)$ which
map states to actions taken in these states~\cite{puterman1994markov,Filar97}.

A major problem with moving from classical planning to MDPs is
representational: in MDPs the full state space $S$ is explicitly
represented, whereas it is only implicit in classical planning, a
result of joint values of state literals.
\emph{Factored MDPs (FMDPs)} aim to address this gap.
Specifically, instead of specifying the state space $S$, FMDPs are represented by a collection $X$ of state
variables (which we take to be binary henceforth); these are random
variables, with $X=x$ denoting a particular instantiation of these to
values $x$.
Dynamics are now represented by a Dynamic Bayes Network for each
action $a$; further, we assume that the reward function is additive in
state variables, i.e., $r(x,a) = \sum_j r_j(x_j,a)$, where $r_j$ are
the variable-specific rewards.
Well-known approaches exist for approximately solving FMDPs~\cite{guestrin2003efficient}.

%% file: planinterdiction.tex
\section{Plan Interdiction Problem}

Our ultimate goal is not merely to compute an optimal plan for the
attacker, but rather to compute \emph{an optimal defender
  interdiction strategy}.
To this end, we model the interaction between the defender and
attacker as a Stackelberg game in which the defender moves first,
choosing to deploy a set of mitigations, and the attacker responds to
these by constructing an optimal attack plan given the resulting
environment.
We can formalize this game as a \emph{plan interdiction problem
  (PIP)}, defined by $\{ \mathcal{M}, c_m, r^{D},\phi\}$, where
$\mathcal{M}$ is the set of defender mitigation actions, $c_m$ is the
cost of a mitigation $m \in \mathcal{M}$, the defender's reward function $r^D(x) = \sum_jr_j(x_j)$, which is additive over state
variables, and the underlying planning problem $\phi$ for the
attacker.
The consequence of a mitigation $m$ can be two-fold: it can modify
current (initial) state $s_0$, and remove a subset of attacker
actions.
Thus, if $S$ is a subset of mitigations used by the defender, these
modify the attacker's planning problem $\phi$; we denote the resulting
modified problem by $\phi(S)$, and the associated optimal plan for the
attacker by $\pi(S)$.
In the PIP, the defender's goal is to choose the optimal set of
mitigations $S$, balancing the defender's utility $V^D(S; \pi(S))$
(total expected discounted reward) and cost of mitigations $c(S) = \sum_{m \in
  S} c_m$:
\begin{equation}
\label{E:pip}
\max_{S \subseteq \mathcal{M}} V^D(S; \pi(S)) - c(S),
\end{equation}
where $\pi(S)$ is the attacker's best response plan, that is, its
optimal plan over the restricted planning domain $\phi(S)$.

We first discuss the special case of this problem in the context of
deterministic planning, and follow with a more general treatment when
uncertainty is involved.

\subsection{Interdiction of Deterministic Plans}

In the deterministic setting, plan interdiction is focused on goals
that can be achieved by the attacker.
Recall that $G$ is the set of goal literals, and $r_l$ is the reward
to the attacker for achieving a goal literal $l \in G$.
We now augment this with a defender's corresponding reward $r_l^D$
(presumably, negative).
The plan interdiction problem then becomes
\[
\max_{S \subseteq \mathcal{M}} \sum_{l \in G} z_l(\pi(S))r_l^D - c(S),
\]
where $z_l$ is a binary indicator of whether the goal literal $l$ is
achieved by the attacker.
Letchford and Vorobeychik \cite{Letchford13} show that this bi-level optimization problem can
be solved using a combination of mixed-integer linear programming and
constraint generation, where constraints represent possible attack
plans, and are iteratively added by computing approximately optimal
plans using state-of-the-art heuristic planning software, such as SGPlan~\cite{Chen06:Temporal}.

\subsection{Interdicting MDPs}

The power of deterministic planning and associated interdiction is
that it involves solving highly structured problems, and we can
therefore develop a scalable approach for these.
The limitation is generality: often, uncertainty in the planning
problem faced by the attacker must be appropriately captured.

We model the attacker's planning problem when it faces uncertainty by an
MDP, described above.
Specifically, we can address the MDP interdiction problem, which can
be defined just as its deterministic counterpart above, but using the
more general utility function $V^D(S; \pi(S))$.
The interdiction problem can then be described very generally by the
Equation~\eqref{E:pip}.

While it is possible to solve MDP interdiction problems using a
similar approach for bi-linear optimization as developed by
Letchford and Vorobeychik~\cite{Letchford13}, these require that the state space be explicitly
represented.
Panda and Vorobeychik \cite{Panda17} addressed this technical challenge by integrating
state-of-the-art approaches for approximately solving FMDPs by
Guestrin et al. \cite{guestrin2003efficient} with a bi-level framework, and appealing
to Fourier representation of a value function over a Boolean
space~\cite{o2008some}.

While the approach by Panda and Vorobeychik \cite{Panda17} enables a considerable advance
in solving MDP interdiction problems over factored MDP
representations, it is still somewhat limited in scalability.
Moreover, a major challenge in plan interdiction approaches to date is
that the defender is typically uncertain about the attacker's planning
problem, such as the initial vulnerabilities and attacker's access and
capabilities, and the resulting Bayesian Stackelberg game is
infeasible for current methods for even small problem instances~\cite{Letchford13}.

In recent research, we have taken an alternative approach which
integrates reinforcement learning with function approximation.
Combined with a restriction that the defender's mitigations only
modify initial state (which is without loss of generality, since we
can capture action removal by adding associated preconditions to
initial state), we can now learn a general value function for the
attacker over the state space, and then solve the interdiction
problem once, without having to iteratively re-solve the planning
problem.
Additionally, we can introduce uncertainty about the attacker by
capturing it as uncertainty over a subset of initial state variables.
In the special case when the value function is linear (or approximated by a linear
function), we can transform the Bayesian interdiction problem into a form which can be solved using
integer linear programming, a significant advance in scalability
compared to prior art.

\subsection{Interdicting Partially Observable MDPs}

MDPs model an attacker's uncertainty about dynamics.
However, they ignore another crucial consideration: attacker's
uncertainty about the true state of the system.
Insofar as this introduces information asymmetry between the defender
and attacker, it introduces a potentially very rich space of
interdiction options for the defender.
One issue that arises is signaling: if the defender knows aspects of
the initial state which are unknown to the attacker, the defender's
mitigations can signal to the attacker information about the true
initial state.
This, in turn, provides an opportunity for deception: the defender
may, through particular (costly) mitigations they deploy signal to an
attacker that they have vulnerabilities they do not actually have,
thereby deceiving the attacker into expending resources into an attack
which cannot succeed.
In addition, the defender may also control the observations of the
attacker about system state, a capability which can provide further
leverage.

As we can see, this \emph{partially observable Markov decision process (POMDP) interdiction problem} is exceptionally rich.
It is also an open problem from a computational perspective---we would
argue, the next important open problem in plan interdiction.

%% file: illustration.tex
\section{Illustration: Threat Risk Assessment Using Plan Interdiction}

Having motivated the plan interdiction problem rather abstractly, we
now illustrate its value more concretely by using the planning
framework for network cyber-risk analysis and mitigation.
Specifically, we first define a model for cyber threat assessment
using classical planning primitives, augmenting these with stochastic
information which represents uncertainty about attacker's access, such
as which user accounts the attacker has already compromised.
We then describe an implementation of this model, and
demonstrate its value through experiments.
Finally, we show that by selectively patching a small set of
vulnerabilities we can dramatically reduce cyber risk.

\input{model}
\input{implementation}
\input{experimentalmethods}
\input{experiments}

\input{threatmin}

%% file: model.tex
\subsection{Model} \label{modeling}
In creating our network threat assessment model, we define six components: a set of hosts $H$, the system environment $T$, a dictionary of vulnerabilities $V$, the target file $F$, and the attacker $A$ with action set $S$. 
%This section of the work describes and 
Next, we formally describe these components.
For illustration, we assume that the attacker aims to exfiltrate a particular file, which may represent sensitive information such as trade secrets.

Each host in our model represents a device within the target environment, with no inherent distinction made among personal computers, servers, mobile devices, etc. Formally, we define a host $h_i \in H$ as possessing three attributes: a set of exploitable vulnerabilities $V_i \subset V$, the set of neighboring hosts $N_i \subset H$, and a flag $F_i \in \{0, 1\}$, which indicates if the target file is present on the host. Vulnerabilities are assigned to a host stochastically, reflecting differences in particular host configurations, non-deterministic exploitation outcomes, and the relative complexity of executing the exploit. The set of neighbors
%is with respect to directed edges within the network; specifically, $N_i$ consists of the hosts to which $h_i$ has an outgoing edge. Put more directly, 
$N_i$ of a host $h_i$ is the set of all hosts which can potentially be accessed from $h_i$.  

The system environment variable $T$ contains the state of each host $t_i$. The state is represented by a binary vector, and contains information pertaining to the status of the host relative to the attacker. For example, whether the attacker has read or write access to files on the host could be a pair of binary values in the state vector.  

Vulnerabilities in the dictionary $V$ are defined by two vectors $R_j, E_j \in \mathcal{R}^E$. The first of these, $R_j$, is the exploitation requirement vector: the minimum required state of the host (relative to the system environment vector for the host) for successful exploitation. One example of a state requirement would be a vulnerability which requires authentication with user credentials. After successful exploitation of a vulnerability, the state of the host is updated according to the exploitation effects vector, $E_j$.

The target file $F$ is simply a binary valued variable that is initially set to 0. If the file is accessed by an attacker, it is updated to 1.

The attacker $A$ 
%serves as the sole agent in our model, and 
possesses three attributes. The first of these is a set of hosts $C \subset H$ for which the attacker possesses user credentials (obtained, for example, through a phishing attack). The other two attributes are also sets of hosts, $H_a \subset H$ and $H_c \subset H$, which are those hosts which are accessible to the attacker and have been compromised by the attacker, respectively. In addition to the aforementioned attributes, the attacker has an abstract set of knowledge and tools to assess information about the system environment (e.g. connectivity between hosts). Further, the attacker is assumed to be able to exploit any vulnerability present on an accessible host.  

The action set of the attacker, $S$, contains five categories of deterministic actions encapsulating reconnaissance, exploitation, and data access/exfiltration. These action categories are summarized formally in Table \ref{table:actions}. The first two categories of actions, Exploration and Probing, deal with efforts by the attacker to learn information about hosts and the network. Exploration targets a single host $h_i$ (which must in the set of compromised hosts) and results in the union of $N_i$, the host's neighborhood, with $H_a$, the set of accessible hosts. Probing examines an accessible host and checks for the presence of particular vulnerabilities.

\begin{table}
\begin{tabular}{|c|c|c|}
	\hline
	Preconditions & Action & Effects \\
	\hline
	$h_i \in H_c$ & $Explore(h_i)$ & $H_a = H_a \bigcup \{N_i\}$ \\
	\hline
	$h_i \in H_a$ & $Probe(h_i, V)$ & $\{v_j \in V | R_j - (R_j \wedge t_i) = 0\}$ \\
	\hline
	$h_i \in H_a \wedge h_i \in C$ & $Masquerade(h_i)$ & $r_i, w_i = 1$ \\
	\hline
	$h_i \in H_a \wedge (R_j - (R_j \wedge t_i) = 0)$ & $Exploit(h_i, v_j)$ & $t_i = t_i \vee E_j$ \\
	%\multirow{2}{55pt}{$h_i \in H_a \wedge (R_j - (R_j \wedge t_i) = 0)$} & \multirow{2}{*}{$Exploit(h_i, v_j)$} & \multirow{2}{*}{$t_i = t_i \vee E_j$} \\ \\
	\hline
	$h_i \in H_c \wedge r_i = 1$ & $Access(h_i)$ & $F = 1$ \\
	\hline
\end{tabular}
\caption{Categories of attacker actions}
\label{table:actions}
\end{table}

The next two categories of actions pertain to malicious behavior executed by the attacker. Masquerading, the first of these actions, is the act of accessing a host through normal means (e.g. SSH) with the use of externally acquired user credentials. Naturally, this requires both that the attacker possesses suitable credentials and access to the host. The result of Masquerading is attainment of read and write privileges (denoted $r_i$ and $w_i$ for host $h_i$) to the target host. Exploitation, the second malicious action available to the attacker, targets a particular accessible host $h_i$ with a vulnerability $v_j$. Satisfaction of vulnerability requirements $R_j$ are necessary for execution of this action, and the effects of exploitation are according to $E_j$. 
The final action available to the attacker is Access. This action targets a particular compromised host $h_i$ and checks for the presence of the target file. If the file is present and the state of the host permits, the attacker accesses the file. Successful execution of this action is the goal for the attacker, and indicates that the system has been compromised successfully. 

%% file: implementation.tex
\subsection{Implementing the Model} \label{impl}

In this section we describe our implementation of the model given in Section \ref{modeling}.

\subsubsection{Vulnerability Dictionary} \label{vulndict}

To populate our dictionary of vulnerabilities, we turn to the National Vulnerability Database (NVD). Originally created in 2000 by the NIST Computer Security Division, the NVD is an extension of MITRE's Common Vulnerability and Exposures (CVE) dictionary that includes additional vulnerability assessment metrics. One of these metrics, the Common Vulnerability Scoring System (CVSS), is of particular interest to us.

Designed to communicate the characteristics and impact of vulnerabilities, the CVSS provides both quantitative scores and vectors of characteristics. The former indicates the holistic ``severity'' of the vulnerability, while the latter gives us some indication of the conditions necessary for and the results of successful exploitation. Three sets of metrics comprise the CVSS: base, temporal, and environmental metrics. Of these we only use the base metrics, which cover intrinsic qualities of a vulnerability. Base metrics are divided into exploitability metrics and impact metrics. The former captures information concerning vulnerability accessibility and has three parameters: attack vector (AV), access complexity (AC), and authentication (Au). Impact metrics gauge the potential effects of successful exploitation, divided into confidentiality, integrity, and availability impacts. As detailed in the next section, each of these parameters contributes to our vulnerability model.

%Temporal score metrics, which pertain to vulnerability characteristics which change over time, have three parameters: exploitability (E), remediation level (RL), and report confidence. Though inclusion of these parameters would enhance our vulnerability model, they are not at present available for many NVD entries. As such, we do not incorporate them.

%Environmental metrics are designed to allow IT personnel to customize the weight given to particular vulnerability characteristics based on the requirements of the organization. This weight customization is effected by modifying the two general modifiers (collateral damage potential and target distribution) and three impact subscore modifiers (confidentiality, integrity, and availability requirements). For example, an organization which routinely handles highly sensitive data may set confidentiality requirements to "high." The result of this modification would increase the severity score for vulnerabilities having confidentiality impacts. As with temporal score metrics, we do not at present include environmental metrics in our vulnerability model.

\subsubsection{Vulnerability Profiles} \label{vulnprof}
Having parsed the NVD to generate a vulnerability dictionary, we next construct profiles containing a subset of these vulnerabilities. 
%Construction can be as simple as randomly sampling the dictionary for a set number of entries or including each vulnerability in a profile with some probability; however, we would like to guide the construction of vulnerability profiles to reflect realistic conditions. 
To this end, we utilize the penetration testing suite Nessus. Specifically, we construct profiles by filtering Nessus vulnerability reports to recover CVE entry IDs, then populate the entries from our dictionary. 

Based on CVSS information from the dictionary, we generate the set of exploitation requirements for each vulnerability and derive a value denoting the exploitation probability. Two base metric parameters are retained for later use: attack vector and authentication. The attack vector can take on one of three values, each of which introduces different preconditions for successful exploitation in our model. The most restrictive value, \textit{Local}, requires an attacker to have either direct physical access to the machine or access to a shell account. A corresponding parameter for each host determines whether this level of access is present. \textit{Adjacent Network} access, the second AV value, necessitates some kind of access to either the broadcast or collision domain of the host. In our model, this access is represented by directed edges between an already compromised system and the target system. Finally, \textit{Network} access indicates that the vulnerability is remotely exploitable. Satisfying \textit{Adjacent Access} or the corresponding network access parameter are required for exploitation in this case.

Authentication requirements can also take on one of three values: \textit{None}, \textit{Single}, and \textit{Multiple}. As the latter two values indicate user credentials must be obtained, they introduce an additional precondition for exploitation of the vulnerability.

Exploitation probability is derived from a combination of base metrics and their corresponding weights in the CVSS severity calculation. The initial value of $1.0$ is first multiplied by the scoring value associated with access complexity. This attribute captures the relative level of exploitation difficulty due to the presence of specialized access conditions (e.g. non-default software configurations). Low complexity, indicative of no specialized access conditions, has an associated value of $0.71$. A value of $0.61$ is used for medium complexity -- so denoted when "somewhat" specialized access conditions are present. High complexity carries an associated value of $0.35$. 
In the case that multiple authentications are required for exploitation of the vulnerability, the probability is multiplied by an additional factor of $0.8$. Finally, we account for the non-deterministic quality of access granted by partial confidentiality impact with a multiplied factor of $0.5$. Note that this value is not derived directly from the CVSS formula.

\paragraph{Phishing Attack Probability} %TODO: cite symantec
Recognizing that attackers are not limited to exploiting software vulnerabilities, we include a phishing attack probability parameter $P_P$ in our framework. Despite the name, the parameter is meant to encapsulate all social engineering vectors which result in the recovery of a user's credentials. In selecting a default value of $0.03$, we examined the spear-phishing rate for small businesses ($<$250 employees) and the overall phishing rate from the 2016 Symantec Internet Security Threat Report. 

In precisely defining the utilization of $P_P$ in our model, we say that it is the probability a given host has one or more privileged user accounts whose credentials have been compromised. In so defining the parameter, we avoid the need to generate a set of user credentials and associate them in some manner with the hosts.  

\paragraph{Zero-Day Attack Probability}
In addition to published vulnerabilities and social engineering vectors, we consider that the attacker may have knowledge of one or more zero-day vulnerabilities which affect hosts.
Since the NVD serves as our repository of vulnerabilities, we utilize it to determine our default probability of zero-day attacks. Specifically, we examined all of the CVE entries between 2002 and 2016 and compared the labeled year of the vulnerability (i.e. the year included in the CVE name) with the published date. If the two differed, the vulnerability was considered to have been present and unknown sufficiently long to qualify as a viable zero-day attack. On average across the span of years mentioned, approximately $13\%$ of vulnerabilities meet this criterion.

\subsubsection{Host Generation} \label{hostgen}
In general, a host in our model can be any device connected to the network; however, for the initial implementation of the framework we limit ourselves to consideration of servers and personal computers. Hosts are defined fundamentally by a set of three attributes: the vulnerability profile, a list of neighboring hosts, and a set of access levels. A unique host name as well as a boolean value which indicates whether the host serves as a gateway are also included. The gateway indicator is used in our generative network model discussed in Section \ref{gen-topo}.

Upon initialization of a host object, the set of vulnerabilities in the profile are iterated through, with each vulnerability being retained or discarded stochastically according to the associated probability of exploitation. This results in each host containing a subset of the full profile and reflects differences in host vulnerability due to varying software configurations.

Neighboring hosts in our model are any hosts which satisfy the properties of adjacent access (as defined in the CVSS). Viewing a particular instance of the model as a graph, each neighbor possesses an incoming directed edge.

The access levels list contains information on host properties related to exploit preconditions. In our initial implementation this consists of three values: network, root, and user access. The first indicates that the attacker can execute remote exploits on the host. Root access implies privilege escalation for the attacker on the host has been successfully carried out. Finally, user access indicates that the attacker possesses either user credentials for an account on the host or the means of equivalently authenticating.

\subsubsection{Generation of PDDL}
 
 The planning domain and problem description for a particular instance are generated sequentially based on instances of the model described in Section \ref{modeling}. First the domain types and constants are produced. The former consists of three entries: \textit{hosts}, \textit{vulnerabilities}, and \textit{files}. The numbered list of hosts and a single generic file are produced as constants. 

Finally, the three actions are constructed. The first of these, the \textit{exploit} action, generalizes the impact of exploiting present vulnerabilities with the intent of gaining read access to the host (to check for the desired file) and, upon failure to locate the file, compromise a neighbor. The base precondition for the action is the presence of a particular vulnerability on the host. Recall that this presence is probabilistic based on the characteristics of the vulnerability and, in the case of heterogeneous host profiles, the vulnerability profile itself. 
Further, we introduce
%Additional preconditions can be introduced again based on the specific vulnerability. Specifically, 
two additional sets of preconditions that can be added in association with specific vulnerabilities. The first is a requirement of user access, and appears in the event that authentication (single or multiple) is required for exploitation of the vulnerability. The second pertains to the required access vector between the attacker and the host: local, adjacent, or network access. Execution of this action grants read access to the attacker and marks the host as compromised. For each vulnerability that appears on at least one host, an instance of this action is added to the domain.

%new stuff

\begin{figure}
	\centering
	\includegraphics[width=1\linewidth]{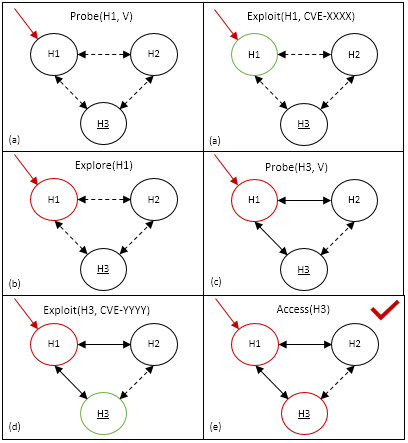}
	\caption{An attack planning example on a small network.}
	\label{fig:planing-cons}
\end{figure}

Figure \ref{fig:planing-cons} depicts a sample plan and the effects of each action for a small example network. In the figure, the red arrow indicates that H1 is remotely accessible, and the underlined text for H3 indicates that the target file is present on the host. Initially, the attacker can only probe for vulnerabilities on H1. Finding a suitable vulnerability, the attacker next compromises H1 by exploiting the flaw. Having penetrated the network, the attacker can explore to discover the connections between H1 and the other hosts. As before, the attacker then probes for and then exploits a vulnerability, this time on H3. Finally, having gained sufficient privileges on the target host, the attacker accesses the target file, indicating the goal condition has been met. 

%Though we model the problem as one where the agent (the attacker) has incomplete information about the system, the implementation is handled as a classical planning problem with complete information. Since the attacker has the means of exploring the network and probing hosts for vulnerabilities, discovery of a path to the target file (if one exists) is inevitable. As we are interested only in whether the file can be reached (not the path taken or its length), it is unnecessary to explicitly implement the probe and explore actions.

%end new stuff

Once a host has been compromised, the number of types of access vectors available to the attacker must be updated accordingly. To this end, we introduced an artificial action, \textit{update access}, which has two preconditions: a connection between two hosts, and that one of the hosts has been previously compromised. Note that connections between hosts are directed. Once these conditions have been met, the neighboring host is updated to reflect that the attacker now has both network and adjacent access vectors.

The third action fulfills the goal condition of the plan, namely, reading the target file. Execution of the \textit{Access} action requires that the attacker have read access to the host which contains the target file. Taking this action indicates that the file has been read successfully, which satisfies the goal conditions of the plan.

Initial generation of the problem instance follows a straightforward process. First, the target file is randomly assigned to one of the hosts. Next, each host is examined and corresponding predicates are added to the problem definition. An instance of the \textit{connected} predicate is generated for each neighbor (i.e. a host that is accessible from this host) based on the state of neighboring hosts. Instances of \textit{has\_vuln} are taken from the specific vulnerability profile of the host. Finally, the presence of initial network, user, or root access are added accordingly. As the final step of problem instance generation, the goal condition of accessing the target file is added.

%% file: experimentalmethods.tex
\subsection{Experimental Methods} \label{methods}

We illustrate our model of cybersecurity risk assessment using a novel experimental framework which combines a generative model of a network with stochastic generation of hosts based on host images. 
%Further, we examine the impact of threat minimization through optimal removal of permissible host connections and patching of vulnerabilities.

\subsubsection{Generating Vulnerability Profiles} \label{gen-profiles}
For our experiments we carefully crafted four vulnerability profiles. Two of these profiles represent generic desktop system configurations, with the remaining two representing server configurations. The operating systems for these profiles were chosen based on compiled OS market share data~\cite{netmarket,stackoverflow}, and software configurations were informed by application use data from WhatPulse~\cite{whatpulse}, a computing habit information aggregator with more than 285,000 active users. 

The first desktop configuration was based on a generic Windows desktop setup. As Windows 7 holds the largest share amongst Windows distributions, it was chosen for use with this profile. In addition to the standard software packages installed with Windows 7, the following pieces of third party software were installed:

\begin{itemize}
	\item Chrome
	\item Firefox
	\item Skype
	\item Java Runtime Environment (JRE)
	\item Adobe Reader
	\item Adobe Flash Player
\end{itemize}

Though Mac OS X holds the second highest market share after Windows for desktops, test images were not readily available for use. This being the case, the second desktop profile was constructed based on Ubuntu Linux version 14.04 (Trusty Tahr). 

For server profiles, we again provided both a Windows and Linux variants. The Windows server is a stock image of Windows Server 2012, while the Linux server is again Ubuntu 14.04, but with the addition of software necessary to run an Apache web server.

\subsubsection{Generating Network Architecture}

We implemented two generative models for constructing random network topologies in the experiments we report on below. The first is the well-known random graph (Erdos-Renyi) model, while the second is a novel generative model aimed at capturing aspects of networks salient in threat modeling.
%of these are drawn from existing techniques, and the third is a novel generative model designed to capture network policy decisions. 

\paragraph{Erdos-Renyi} \label{er-topo}
The first model implemented was the Erdos-Renyi random graph model \cite{erdos1960}. 
Specifically, we utilize the $G(n, p)$ formulation of the model in which a random graph $G$ is constructed by randomly connecting vertices. For every pair of vertices $v_1, v_2 \in G$, an edge $e$ is added between them with probability $p$, independently of any other edge. 

In addition to specifying arbitrary values of $p$, we can utilize relationships and established thresholds between $n$ and $p$ to construct graphs with known, potentially interesting characteristics. An example of a known threshold in Erdos-Renyi graphs is $p = ln(n) / n$. This value represents a sharp connectedness threshold -- larger values of $p$ will almost certainly be connected, while smaller values indicate the almost sure certain presence of isolated vertices in the graph. Several of these thresholds are employed in the experimental evaluations detailed in \ref{er-exp}. 

\paragraph{Organizational Network Generative Model} \label{gen-topo}

The Erdos-Renyi model offers clean insights, but is known to be overly simplistic to realistic topological characteristics of real networks.
Moreover, in modeling security threats on networks one must also make a distinction between types of hosts (such as servers and desktops), as that pertains to the particular OS that is likely to run on such systems (for example, servers will typically not run Microsoft Word, and would not be susceptible to vulnerabilities in this application).
To this end, we use a more realistic Block Two-Level Erdos-Renyi (BTER) model proposed in \cite{sesh2012}, and extend it with host-level characteristics.

%General graph models provide a base for the generation of network topologies in our framework; however, it is desirable to have a means of generating topologies which more closely model computer networks specifically and permit procedural classification of devices as well as definable connectivity policies. 
%To this end, we introduce a generative model which incorporates portions of a 
%BTER generation process. 

%In the interest of better capturing the characteristics of real networks, we also incorporate into the framework the ability to generate topologies based on the Block Two-Level Erdos-Renyi (BTER) model proposed in \cite{sesh2012}. 
Unlike the original Erdos-Renyi model, which treats all vertices identically with regard to edge construction, BTER treats the overall graph structure as a set of interconnected communities, each of which is an Erdos-Renyi random graph. 
The first phase of BTER graph construction consists of generating a collection of Erdos-Renyi (ER) blocks according to a user-specified degree distribution. These blocks are interconnected in the second phase via nodes in each block that have access degree. The Chung-Lu graph model \cite{chunglu2001} (which can be framed as a weighted Erdos-Renyi graph) is employed over the excess degrees to form the connections among the blocks.

\begin{figure}
	\centering
	\includegraphics[width=1\linewidth]{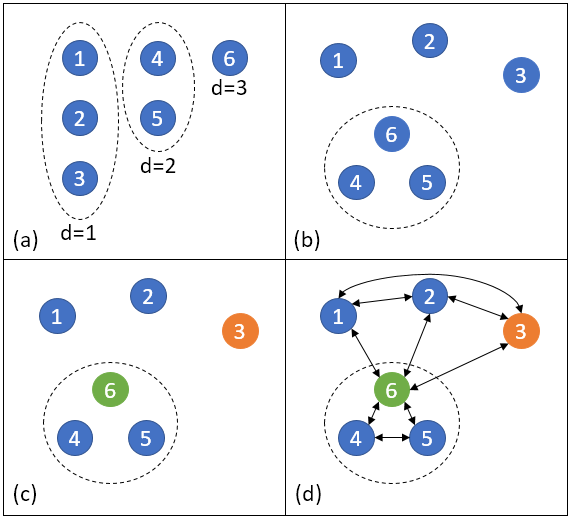}
	\caption{Generative Model Network Construction}
	\label{fig:gen-model-cons}
\end{figure}

Our model takes as input a desired node count $n$, a power law exponent $\alpha$, a remote access probability $P_N$, and an optional set of connectivity policies $\rho$. 
%Per \cite{barabasi1999}, it is recommended that values between 2.1 and 4 be used for $\alpha$.
Construction of a network with our model then follows four primary steps, the explanation of which will be aided by Figure \ref{fig:gen-model-cons}. We first sample $n$ discrete values from a power law sequence with exponent $\alpha$. In the BTER construction process, these values represent the desired degree of the nodes, and define the size of the communities present within the graph. The sampled values are sorted in ascending order and grouped by value, as seen in panel (a) of Figure \ref{fig:gen-model-cons}. 
Each entry in the sample list with value 1 represents a device which does not belong to a restricted subnet. As such, we don't consider them when building communities. Starting with the first value greater than 1, groups are formed by selecting and removing the first $d + 1$ entries from the list, where $d$ is the value of the first entry. This procedure continues until the list is empty, at which point all of the groups have been formed. The end result is displayed in panel (b) of Figure \ref{fig:gen-model-cons}: a set of individual nodes and one or more communities of nodes.   

Device classification occurs next, with two sets of distinctions made among the devices. First, one node in each community is chosen arbitrarily to serve as the gateway node. Next, each individual node (i.e. those not in communities) is flagged as remotely accessible with probability $P_N$; those flagged are distinguished as server nodes. In panel (c) of Figure \ref{fig:gen-model-cons}, node 6 is a gateway and node 3 is a server.

Finally, edges are drawn between the nodes as shown in panel (d). By default, all individual nodes are bi-directionally adjacent to each other and the set of gateway nodes. Additionally, each community forms a complete subgraph. As shown later in Section \ref{gen-exp}, the policies defined in $\rho$ modify the manner in which edges are added.

%% file: experiments.tex
\subsection{Experimental Study of Network Cyber Risk}

We now present an experimental study of network cyber risk using the experimental setup in Section~\ref{methods} which instantiates our planning-based threat analysis framework.

\subsubsection{Erdos-Renyi Network Model} \label{er-exp}
As an initial set of experiments, we employ the Erdos-Renyi random graph model with directed edges to generate our network topology. As mentioned in Section \ref{er-topo}, graphs generated using the Erdos-Renyi model possess well-known characteristics based on the relationship between the number of hosts $n$ and the edge probability $p$. We examine thresholds corresponding to two of these relationships: 

\begin{enumerate}
	\item $np = 1$
	\item $p = \frac{ln(n)}{n}$
\end{enumerate}

The first of these, which we will henceforth refer to as $NP1$, produces graphs which almost always contain a maximum component of size $n^{\frac{2}{3}}$. The second relationship, referred to hereafter as $LNN$, is the threshold for connectedness in the model \cite{erdos1960}. 
%For the evaluations shown in Figure \ref{fig:er_1}, the values of $p$ are therefore $0.01$ and $0.046$ for $NP!$ and $LNN$, respectively.
Since the Erdos-Renyi model does not lend itself to differentiation of vertices based on topological characteristics, we generate hosts homogeneously using the Windows desktop vulnerability profile detailed in Section \ref{gen-profiles}. 

\begin{figure}
\centering
\includegraphics[width=1\linewidth]{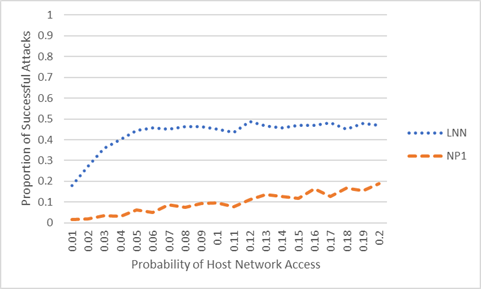}
\caption{Cyber risk for Erdos-Renyi Topologies, $n = 100$.}
\label{fig:er_1}
\end{figure}

Figure \ref{fig:er_1} shows the effect of varying the host network access probability $P_N$ for a range of values between $0.01$ and $0.20$. As we might expect, the behavior for the two different values of $p$ differ significantly. For the $NP1$ case, the proportion of successful attacks $P_S$ is much smaller, but increases roughly linearly with $P_N$ (indeed, across the range of values we considered, $P_S \approx P_N$ in this case). 
%Further, the proportion of successes is quite low -- across the range of values tested, $P_S$ is approximately equal to $P_N$. 
We can explain this behavior by examining the structure of the generated networks. Since the attacker does not have the option of physically accessing devices, attack vectors must originate with devices that are remotely accessible. Sparsity of edges, their directed nature, and the inherently disconnected structure of the generated networks limits the ability of the attacker to penetrate further into the network following the compromise of an externally accessible device. Further, since zero-day attacks and utilization of phished user credentials require network access to vulnerable devices, they are of relatively little use for this topology.

With the $LNN$ case, the effect of varying $P_N$ is completely different. Following a sharp increase in $P_S$ across $P_N$ values between $0.01$ and $0.05$, the proportion of successful attacks reaches a steady state. Given the guarantee of connectedness inherent to the $LNN$ case, any remotely-accessible device will have a path to the target device; however, no guarantees are made that any two remotely-accessible devices will have independent paths to the target. The leveling-off of $P_S$ suggests that independent paths to the target device are saturated beyond $P_N = 0.05$.  

\subsubsection{Organizational Network Generative Model} \label{gen-exp}
As a follow-up to the initial experiments run with the Erdos-Renyi model, similar evaluations with the generative model proposed in Section \ref{gen-topo} are executed. Except where otherwise noted, default values for $P_P$ and the zero-day attack probability $P_Z$ are used (0.03 and 0.13, respectively). For the generative model, we use directed edges and also differentiate vulnerability profiles across the hosts. Gateway and server nodes (remotely accessible nodes not in a community) are assigned either the Linux or Windows server profiles with probabilities $0.664$ and $0.336$, respectively. All other nodes are assigned either the Windows or Linux desktop profiles with probability $0.7048$ and $0.2952$. These values represent the proportional ratio of Windows to Linux according to \cite{netmarket}. In addition to varying the network access probability, $\alpha$ value, and host count, we also examine the impact of several network policies and the effect of increased phishing attack probabilities.
For the baseline version of the generative model, phishing attacks have an additional effect: if at least one host has compromised credentials, then the attacker is able to utilize a VPN connection, resulting in servers, gateways, and non-community desktop nodes being remotely accessible. 

\begin{figure}
	\centering
	\includegraphics[width=1\linewidth]{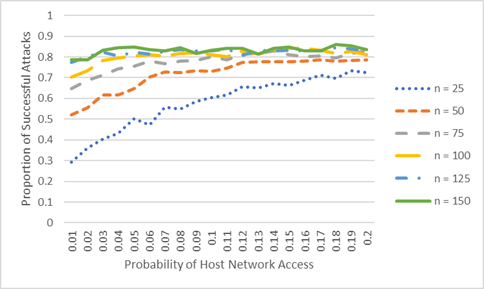}
	\caption{Cyber risk for the organizational network generative model with varying host counts, no connection restrictions. $\alpha = 2.5$}
	\label{fig:gen1-host}
\end{figure}

Figure \ref{fig:gen1-host} demonstrates the impact of varying network size for the baseline organizational network generative model. For small values of $P_N$, there are significant differences between $P_S$ across smaller network sizes. These differences become smaller as the largest network sizes are reached. 
Moreover, risk increases with the number of hosts.
%The difference between host counts is much more pronounced than in the equivalent BTER experiment: 
%the additional effect of successful phishing attack helps explain this difference. 
The intuition behind these observations is that for smaller networks we expect that phishing attacks are less likely (a lower probability that \emph{some} user account is phished), which limits attack vectors to remotely accessible nodes. Larger networks have more remotely accessible hosts, and a higher likelihood that some credentials are compromised through a phishing attack.
Moreover, once even a single credential is phished, the attacker can remotely access a much larger portion of the network, significantly increasing the threat of a successful cyber attack.
Consequently, increasing network size increases overall risk in this model.
%, which results in a much quicker convergence of $P_S$ to a steady state. 

\begin{figure}
	\centering
	\includegraphics[width=1\linewidth]{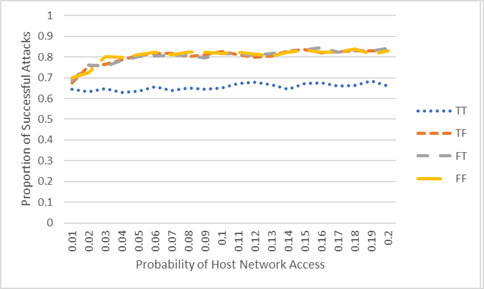}
	\caption{Cyber risk for the organizational network generative model with varying connection restrictions, $n=100$. On the legend, the first letter denotes gateway restrictions (True/False), and the second denotes server restrictions}
	\label{fig:gen1-comp}
\end{figure}

We next examine the effect of introducing two network connection policies. The first of these is a restriction on outgoing connections from gateway nodes. If enabled, gateways retain outgoing connections only to nodes within their subnet, and all other outgoing connections are removed; incoming connections are not modified. The second policy restricts all outgoing connections from server nodes. 
In other words, if the second policy is in place, one cannot connect (e.g., SSH) from a server to any other device on the local network.
Figure \ref{fig:gen1-comp} displays each of the combinations of these two policies. On the legend, the first letter denotes gateway restrictions (True/False), and the second denotes server restrictions. We can see that either of these policies by itself does not appreciably improve the security of the network compared to the baseline.
Surprisingly, however, combining them results in an average reduction of $0.15$ in risk, $P_S$. 
Note that because of the attacker's ability to employ VPN connections, there is little sensitivity to changes in $P_N$.

\begin{figure}
	\centering
	\includegraphics[width=1\linewidth]{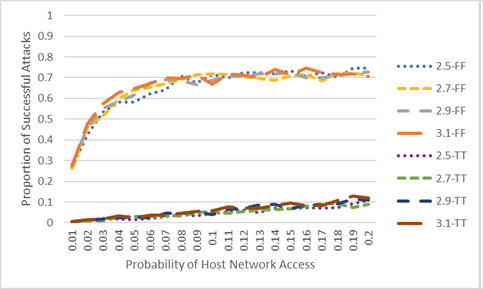}
	\caption{Cyber risk for the organizational network generative model with no VPN. On the legend, the number denotes the value of $\alpha$, 'FF' denotes no access restrictions, and 'TT' denotes restricted gateways and servers.}
	\label{fig:gen1-novpn}
\end{figure}

Next, we study the risk associated with allowing VPN connectivity.
As seen in Figure \ref{fig:gen1-novpn}, disabling VPN connections to the network has a significant impact on overall cyber risk. For the unrestricted model across different values of $\alpha$, we see high initial sensitivity to changes in $P_N$. As the attacker can no longer bypass attacking remotely accessible servers, the expected number of such servers in the network becomes quite relevant. For a network of 100 nodes, $P_N = 0.08$ marks the beginning of the steady state for $P_S$. Restricting both gateways and servers results in extremely small values of $P_S$, and retains only minimal sensitivity to increases in $P_N$.

\begin{figure}
	\centering
	\includegraphics[width=1\linewidth]{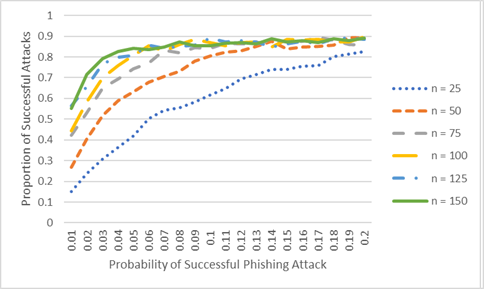}
	\caption{Cyber risk for the organizational network generative model with varying phishing attack probabilities. $\alpha = 2.5$}
	\label{fig:gen1-phish}
\end{figure}

In assessing the impact of increased phishing attacks on the system, we varied $P_P$ from $0.01$ to $0.20$ in increments of $0.01$ while keeping $P_N$ at $0.01$. The results of this experiment are shown in Figure \ref{fig:gen1-phish}. As we might expect, both the initial values of $P_S$ and the sensitivity to increases in $P_P$ are markedly higher in larger networks. We've seen that the ability to bypass attack bottlenecks on remotely accessible hosts by utilizing VPN connections has a profound impact on security, and these results reinforce this observation. Even the network of 25 hosts approaches the same level of vulnerability as the 150 host network when $P_P$ becomes large.

%% file: threatmin.tex
\subsection{Cyber Risk Mitigation through Plan Interdiction}
%Beyond assessing targeted risk for topologies coupled with varying model parameters, threat minimization is a useful tool for systems administrators seeking to ameliorate weaknesses in their networks. Utilizing a greedy approach with respect to overall system vulnerability, an exploration of threat minimization through two means is considered: removal of connections between hosts (graph edges), and removal of known vulnerabilities. 

Our discussion of the risk assessment approach, which leverages AI planning, thus far focused on the attacker modeling.
Although experiments demonstrated the value of specific policy changes in mitigating some cyber risk, we only considered a small set of such policies.
We now explore systematically a common challenge faced by organizations: of the many vulnerabilities present on the network, which should be prioritized.
While in an ideal world, all vulnerabilities are patched, in reality doing so can be both time consuming, and costly in terms of lost productivity.
Consequently, in practice a natural question is whether an organization can focus on a small subset of vulnerabilities which are truly critical for this organization, and make these a high priority.
Given that our risk analysis framework is based on attack planning, the problem of choosing an optimal subset of vulnerabilities to prioritize is precisely the plan interdiction problem.

To formalize, let $V$ be the set of all vulnerabilities present on a network, and let $S \subseteq V$ be a subset of these which the defender (network administrator) will prioritize.
Rather than imposing a cost on patching a given vulnerabilty and minimizing risk, as in our model above, we minimized the number of vulnerabilities that are prioritized, subject to a constraint that risk is below a target threshold $\theta$.

Let $R(S)$ be the risk (probability of a successful attack) faced by the defender when the set $S$ of vulnerabilities is chosen to be prioritized for patching.
Implicitly, $R(S)$ depends on the attacker's policy in response to $S$.
However, note that in our framework, the attacker's policy is also a function of additional exogenously specified factors, such as the assignment of (exploitable) vulnerabilities to hosts, and phishing attack success, among others.
Let $\eta$ be a random variable which captures these stochastic factors, and let $\pi(S,\eta)$ be the associated optimal plan for the attacker.
If we define $r(S,\pi(S,\eta))$ as a binary variable indicating whether the attack succeeds under exogenous parameters $\eta$, risk becomes $R(S) = \mathbb{E}_\eta [r(S,\pi(S,\eta);\eta)]$.
The defender's optimization problem is then
\begin{align}
\label{E:riskmin}
\min_{S \subseteq V} |S| \quad \mathrm{s.t.:} \quad R(S) \le \theta.
\end{align}
We propose a heuristic solution approach for this problem: greedily add a vulnerability $v \in V$ to $S$ in the order of maximum marginal impact on reducing risk $R(S)$, until we satisfy the condition $R(S) \le \theta$.

Next, we evaluate the impact that selective patching of vulnerabilities has on cyber risk.
For these evaluations, we set $P_N = 0.01$, and used default values for $P_P$ (probability of a successful phishing attack) and $P_Z$ (zero-day probability). Additionally, $\alpha$ was fixed at $2.5$. We used the organizational network generative model to generate the network topologies, along with the heterogeneous vulnerability profiles and varied host count to assess the impact of vulnerability patching for different network sizes, as in the earlier experiments. 
We set the threshold in the optimization problem~\eqref{E:riskmin} to $\theta = 0.05$. 

\begin{figure}
	\centering
	\includegraphics[width=1\linewidth]{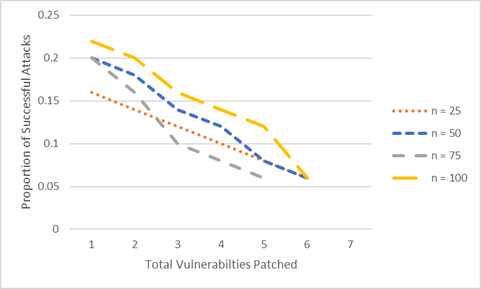}
	\caption{Threat minimization through vulnerability prioritization. On the legend, the value of $n$ denotes the number of hosts in the network.}
	\label{fig:threat-vuln}
\end{figure}
Figure \ref{fig:threat-vuln} shows the results of our heuristic prioritization approach as a function of the number of vulnerabilities being prioritized (until the threshold of $\theta = 0.05$ is reached). As we might expect, the larger networks display higher initial vulnerability; however, the response of each network to the removal of vulnerabilities was very similar in magnitude. Most of the vulnerabilities patched came from the Windows Desktop vulnerability profile, which contributed 81 of the 104 vulnerabilities (after filtering them by year) among all profiles.  
These initially patched vulnerabilities were consistent in that they possessed a higher than average probability of being exploitable on any given host with the profile.
Remarkably, however, we can dramatically reduce cyber attack risk in this model by patching relatively few vulnerabilities ($<10$), as long as we effectively prioritize these.

%% file: dynamicpi.tex
\section{Dynamic Plan Interdiction Games}

Thus far, our discussion (and most prior research) had focused on
interdiction games in which the defender first makes a fixed decision,
and then the attacker devises a dynamic plan, with the possibility of
circumventing defender's mitigations.
This is most relevant when the defender cannot easily observe
adversarial actions as they unfold, and only detects attacks after
they have either succeeded or failed.
In the remainder of this chapter we briefly discuss ideas for modeling
the more general situation in which the
defender may detect portions of the attack plan, and can subsequently
react to it, for example, deploying additional mitigations in response
(including modifications to the observed state).

Let's start with the following scenario: the defender has deployed
mitigations, and a detection system.  
Let's view the detection system as predicting whether or not a
particular observed state $x$ involves malicious activity, captured by
a binary classifier $f(x)$.
Suppose that the detectors are binary, and we can immediately deploy response
mitigations that prevent a detected attack from succeeding.
The
problem of designing such detectors can be viewed as a ``basic'' plan
interdiction problem: we aim to design detectors which minimize loss
from undetected attacks, where mitigation (detector design) costs can
capture the costs of false positives.

An interesting challenge arises when detectors may detect an attack,
but optimal mitigations may be unclear.
This arises when the defender is uncertain about the attacker---for
example, about the attacker's goals.
In this case, the defender may have discovered a subplan (a partial
sequence of attack actions), but is uncertain about the full plan.
This gives rise to an interesting problem which couples plan
interdiction with \emph{plan recognition}: using partial information
about the observed plan to infer information about the attacker's
planning problem, so that mitigations can be optimally deployed.
If the attacker is myopic, this can in principle be addressed through
Bayesian updating: the defender may have a prior over the attacker's
planning problem aspects (such as the attacker's capabilities and
access), and then observations of an actual attack allows them to
infer a posterior distribution over these (potentially resolving much
of the uncertainty).
However, an intelligent attacker may devise plans so as to make such
inference difficult, and faces an interesting tradeoff between
deceiving the defender, and achieving their goals as efficiently as
possible.
We view the challenge of addressing such dynamic problems at a sufficiently high resolution to
obtain practically meaningful results as a major open problem in plan
interdiction games.

%% file: conclusion.tex
\section{Conclusion}

We introduced the general notion of \emph{plan interdiction games}.
In their most basic variation, the defender commits to a collection of
mitigations, and the attacker subsequently chooses an optimal attack
plan.
We illustrate the value of this modeling framework by using it to
develop a network cyber threat assessment approach in which network
features, as well as vulnerability profiles from standard
vulnerability datasets, are used to compute a risk of successful
attacks.
Our experiments offered several interesting insights into the nature
of risk associated with different design choices, such as specific
choices about network connectivity and the use of VPN, as well as
different exogenous factors, such as susceptibility to phishing
attacks.
Moreover, we demonstrated an important use case of plan interdiction
as a means to prioritize which vulnerabilities are immediately
patched, and showed that prioritizing only a few vulnerabilities can
have a significant impact on reducing exposure to cyber risk.

While we have made considerable progress modeling the basic plan
interdiction problem,
as well as advancing technical state of the art to significantly
improve scalability and generality of solution approaches,
a number of interesting conceptual and technical issues arise when uncertainty is
involved either on the part of the attacker about system state, or on
the part of the defender about the attacker's planning problem
primitives.
Dynamics of the problem, in addition to the different aspects of
information asymmetry, allow us to consider a very rich space of
issues of great relevance in cybersecurity, including adaptive defense
and deception both on the part of the defender, and on the part of the
attacker.
A great deal more research is needed to fully understand the space of plan
interdiction problems, develop scalable solutions, and apply these in
practical cybersecurity settings.

%% file: tmplans.bbl
\begin{thebibliography}{10}
\providecommand{\url}[1]{{#1}}
\providecommand{\urlprefix}{URL }
\expandafter\ifx\csname urlstyle\endcsname\relax
  \providecommand{\doi}[1]{DOI~\discretionary{}{}{}#1}\else
  \providecommand{\doi}{DOI~\discretionary{}{}{}\begingroup
  \urlstyle{rm}\Url}\fi

\bibitem{netmarket}
Desktop operating system market share.
\newblock \urlprefix\url{https://www.netmarketshare.com/}

\bibitem{stackoverflow}
Developer survey results 2016.
\newblock
  \urlprefix\url{https://insights.stackoverflow.com/survey/2016#technology-development-environments}

\bibitem{whatpulse}
Whatpulse: Most used applications.
\newblock \urlprefix\url{https://whatpulse.org/stats/apps/}

\bibitem{chunglu2001}
Aiello, W., Chung, F., Lu, L.: A random graph model for power law graphs.
\newblock Experimental Mathematics \textbf{10}(1), 53--66 (2001).
\newblock \urlprefix\url{http://eudml.org/doc/227051}

\bibitem{Chen06:Temporal}
Chen, Y., Wah, B.W., wei Hsu, C.: Temporal planning using subgoal partitioning
  and resolution in {SGPlan}.
\newblock Journal of Artificial Intelligence Research \textbf{26}, 323--369
  (2006)

\bibitem{erdos1960}
Erdos, P., R{\'e}nyi, A.: On the evolution of random graphs.
\newblock Publ. Math. Inst. Hung. Acad. Sci \textbf{5}(1), 17--60 (1960)

\bibitem{Filar97}
Filar, J., Vrieze, K.: Competitive Markov Decision Processes.
\newblock Springer-Verlag (1997)

\bibitem{Ghare71}
Ghare, P., Montgomery, D., Turner, W.: Optimal interdiction policy for a flow
  network.
\newblock Naval Research Logistics Quarterly \textbf{18}(1), 37--45 (1971)

\bibitem{guestrin2003efficient}
Guestrin, C., Koller, D., Parr, R., Venkataraman, S.: Efficient solution
  algorithms for factored mdps.
\newblock Journal of Artificial Intelligence Research \textbf{19}, 399--468
  (2003)

\bibitem{Letchford13}
Letchford, J., Vorobeychik, Y.: Optimal interdiction of attack plans.
\newblock In: International Conference on Autonomous Agents and Multiagent
  Systems, pp. 199--206 (2013)

\bibitem{McMasters70}
McMasters, A., Mustin, T.: Optimal interdiction of a supply network.
\newblock Naval Research Logistics Quarterly \textbf{17}(3), 261--268 (1970)

\bibitem{o2008some}
O'Donnell, R.: Some topics in analysis of boolean functions.
\newblock In: Proceedings of the fortieth annual ACM symposium on Theory of
  computing, pp. 569--578. ACM (2008)

\bibitem{Panda17}
Panda, S., Vorobeychik, Y.: Near-optimal interdiction of factored mdps.
\newblock In: Conference on Uncertainty in Artificial Intelligence (2017)

\bibitem{puterman1994markov}
Puterman, M.L.: Markov Decision Processes: Discrete Stochastic Dynamic
  Programming.
\newblock John Wiley \& Sons, Inc. (1994)

\bibitem{Salmeron09}
Salmeron, J., Wood, K., Baldrick, R.: Worst-case interdiction analysis of
  large-scale electric power grids.
\newblock IEEE Transactions on Power Systems \textbf{24}(1), 96--104 (2009)

\bibitem{sesh2012}
Seshadhri, C., Kolda, T.G., Pinar, A.: Community structure and scale-free
  collections of erd{\H{o}}s-r{\'e}nyi graphs.
\newblock Physical Review E \textbf{85}(5), 056,109 (2012)

\end{thebibliography}
